# Narrow head-tail radio galaxies at very high resolution

B. Terni de Gregory[1], L. Feretti[1], G. Giovannini[1,2], F. Govoni[3], M. Murgia[3], R.A. Perley[4], V. Vacca[3]

[1] Istituto di Radioastronomia INAF, via Gobetti 101, I–40129 Bologna, Italy
[2] DIFA, Università di Bologna, Via Gobetti 93/2, I–40129 Bologna, Italy
[3] Osservatorio Astronomico di Cagliari INAF, Via della Scienza 5, I–09047 Selargius (CA), Italy
[4] National Radio Astronomy Observatory, PO Box 0, Soccoro, NM–87801

**ABSTRACT**

*Aims.* Narrow-angle tailed (NAT) sources in clusters of galaxies can show on the large scale very narrow tails that are unresolved even at arcsecond resolution. These sources could therefore be classified as one-sided jets. The aim of this paper is to gain new insight into the structure of these sources, and establish whether they are genuine one-sided objects, or if they are two-sided sources.
*Methods.* We observed a sample of apparently one-sided NAT sources at subarcsecond resolution to obtain detailed information on their structure in the nuclear regions of radio galaxies.
*Results.* Most radio galaxies are found to show two-sided jets with sharp bends, and therefore the sources are similar to the more classical NATs, which are affected by strong projection effects.

## 1. Introduction

Fanaroff-Riley type I sources (FR I) are low-power radio galaxies ($P_{1.4GHz} < 10^{24.5}$ W/Hz) characterized by jets that are relativistic on the parsec (pc) scale (in agreement with unified models, see, e.g., Urry & Padovani 1995, Giovannini et al. 2001), but the jet velocity slows down and jets become subrelativistic at a few kpc from the core (e.g., Liuzzo et al. 2009). FR I sources present a variety of morphologies. In general, they are characterized by brightness that gradually decreases, a spectrum that steepens toward the lobes, and jets that are characterized by large opening angles. Some of them, however, are more similar to FR II sources, with a spectrum that steepens from the outer edge of the lobe inward (Parma et al. 1999).

Fanaroff-Riley I radio galaxies in clusters show a well-known distorted morphology caused by the interaction of moving galaxies with the intracluster medium (ICM) and are classified as narrow-angle (NAT) and wide-angle (WAT) tailed sources. NATs are common in clusters of galaxies, and their distinctive structures reveal strong interactions with the cluster media. This is driven by transonic or supersonic relative motions. A few NAT sources have been found to show a remarkably narrow extended structure, with the two radio tails not resolved even with high-resolution (arcsec scale) radio observations. Typical examples are IC 310 (see, e.g., Feretti et al. 1998 and references therein) in the Perseus cluster and sources C and I in A 2256 (Miller et al. 2003). Moreover, in these sources no extended emission (radio lobe) is visible at the end of the narrow jet, and the extended morphology of these sources is similar to that of a one-sided naked jet.

The MAGIC telescopes detected very high energy (TeV) gamma-ray emission from IC 310 (Aleksić et al. 2010). By means of VLBA radio observations, Kadler et al. (2012) detected a one-sided pc scale jet at the same position angle of the kpc scale radio structure. They interpreted this radio structure as due to two intrinsically symmetric relativistic radio jets, where only the approaching jet is visible because of Doppler boosting. They estimated that IC 310 must be oriented at a relatively small angle with respect to the line of sight ($\theta < 35°$) and should be classified, in agreement with unified models, as a low-luminosity FRI radio galaxy at a borderline angle to reveal its BL Lac-type central engine. The blazar-like nature of IC 310 is supported by a weak nonthermal activity that is present in optical data and by the detection in X-ray of an unresolved source at the position of the IC 310 nucleus (Kadler et al. 2012, and references therein). The interpretation of IC 310 as a one-sided naked-relativistic jet on such a large scale is puzzling: the one-sided radio structure of IC 310 indeed extends up to a projected distance of 400 kpc. If the one-sidedness is due to relativistic effects, it implies that jets are still collimated and relativistic at a large distance from the core, whereas the jets in low-power sources (FRI and BL Lac) are expected to decelerate on the kpc scale. We note here that the well-known BL Lac MKN501 is one-sided on the pc scale (Doppler-boosting effect), but it shows a symmetric structure on a scale of a few tens of kpc (Giroletti et al. 2004), as well as the bright BL Lac PKS 0521-365, where a bright one-sided jet is present (with emission also in the optical band), but a radio lobe on the other side is visible (Falomo et al. 2009).

To shed light on the structure and nature of narrow NAT sources, we selected sources from published data that are characterized by a very narrow-angle tail structure, transversally unresolved in images at the arcsecond in resolution. The sample consists of 11 sources in 9 clusters of galaxies. Although this is not a complete sample, it is broad enough to allow a study of these sources and obtain their properties, to identify genuine one-sided objects, or sources characterized by symmetric jets with a strong bend near to the core region.

**Table 1.** Angular to linear size conversion with the adopted cosmology

| Name | redshift | conversion factor kpc/″ |
|---|---|---|
| A 278 | 0.089 | 1.6 |
| A 520 | 0.199 | 3.2 |
| A 869 | 0.117 | 2.1 |
| A 1775 | 0.072 | 1.3 |
| A 1795 | 0.063 | 1.2 |
| A 2142 | 0.091 | 1.6 |
| A 2255 | 0.081 | 1.5 |
| A 2256 | 0.058 | 1.1 |
| Ophiuchus | 0.028 | 0.5 |

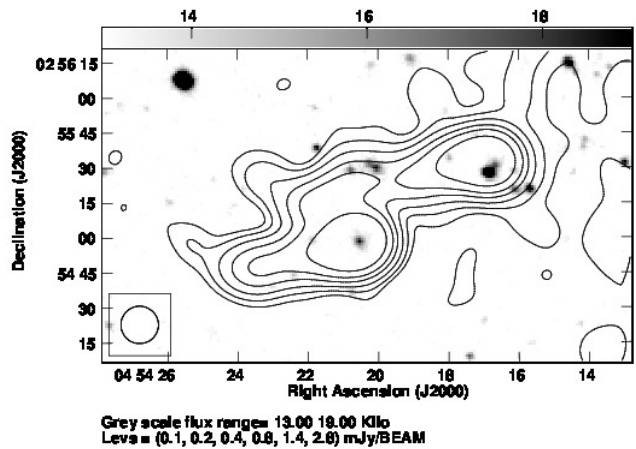

**Fig. 1.** Contour image of A 520 at 1.4 GHz showing the two NAT radio galaxies on the large scale (from Govoni et al. 2001), overlaid on the optical image from the Sloan Digital Sky Survey (SDSS).

In Sect. 2 we present the radio data reduction. In Sect. 3 we give the images and results. The discussion is presented in Sect. 4, and our conclusions are summarized in Sect. 5.

Throughout the paper we use the $\Lambda$CDM cosmology with $H_0 = 71$ km s$^{-1}$Mpc$^{-1}$, $\Omega_m = 0.27$, and $\Omega_\Lambda = 0.73$.

The angular/linear size conversion factor with the adopted cosmology is given in Table 1.

## 2. Radio data

Radio data were obtained with the Karl G. Jansky Very Large Array (JVLA) of the National Radio Astronomy Observatory from May to September 2015, and they are summarized in Table 2. All radio galaxies were observed in the Ku frequency band, and only for some of them, observing time was also granted in the C and L band. The Ku band covers a frequency range from 11.76 GHz to 18.28 GHz. The data are divided into 64 spectral windows of 128 MHz bandwidth (with some windows having the same frequency), and each spectral window is divided into 64 channels of 2 MHz. The frequency of C band ranges from 4.55 GHz to 6.45 GHz and is divided into 16 spectral windows of 128 MHz (with some small overlap). Each window is divided into 64 channels of 2 MHz. The L-band frequency coverage ranges from 0.80 GHz to 2.20 GHz. There are 16 spectral windows of 128 MHz width (with some overlap), each divided into 64 channels of 2 MHz.

The JVLA was in the A configuration for all targets and all observing bands, except for the southern cluster Ophiuchus, which has been observed in the hybrid BnA configuration to obtain a roughly circular restoring beam.

The sources 3C 286 or 3C 138 were observed as calibrators of the flux density scale and of the bandpass, while strong nearby calibrator sources were observed for the calibration of antenna gains and phases.

The data were calibrated and reduced with the Astronomical Image Processing System (AIPS) of the National Radio Astronomy Observatory following the standard procedure. After applying significant editing to the uv data to identify and remove bad data, channels were averaged in each IF. Images were produced by Fourier-transform, Clean and Restore, using the task IMAGR. For each source, images at slightly different angular resolutions were obtained by specifying different weights to the visibilities in the uv plane (ROBUST parameter) or by selecting the uv range. Only in a few cases, that is, in A 869 and Ophiuchus, the source is strong enough to allow self-calibration to minimize the effects of amplitude and phase variations. At each step, clean components were carefully selected as a model for the self-calibration. Gain calibration solutions were obtained with a long integration time only after a few cycles of phase calibration. The process was stopped when no further significant improvement was obtained. Then, a deep cleaning was applied to the data until the total cleaned flux reached a stable value. We are aware that more modern CLEAN algorithms are available both in AIPS and in other packages, however they are not needed here because the targets are of small size and are located at the field center.

We concentrated our study on the total intensity emission, thus we performed the observations without including the calibration of polarization.

## 3. Results

Images produced at 15 GHz have angular resolutions of 0.1″-0.15″. The achieved sensitivities are generally lower than $10^{-2}$ mJy/beam. Some sources are very faint, and the structure is only detected at low levels of significance. The sensitivities of C- and L-band observations are on the order of few $10^{-2}$ mJy/beam, with an angular resolution from 0.3″ to 1″, respectively.

In the following we present images and notes for the individual radio sources and present the results. The precise resolution (FWHM) and noise level of each image is given in the figure caption. The first contour is generally at the $3\sigma$ level (negative and positive), but in a few cases of sources with very weak structure, the first contour is at the $2\sigma$ level. In some cases, gray scales are overimposed for clarity to highlight structure minima. A cross, when present, indicates the position of the optical galaxy.

**Table 2.** Observed radio galaxies

| Cluster | Source name | Other source name | RA (J2000) h m s | DEC ° ′ ″ | Band | Obs time min | Ref |
|---|---|---|---|---|---|---|---|
| A 278 | J0157+3214 | B0154+320 | 01 57 16.0 | +32 14 48.0 | Ku | 26 | 1 |
| A 520 | J0454+0255A | D | 04 54 19.0 | +02 55 10.0 | Ku | 26 | 2 |
| | J0454+0255B | E | | | | | |
| A 869 | J0946+0222 | B0943+026 | 09 46 14.0 | +02 22 47.0 | Ku | 26 | 1 |
| | | | | | C | 16 | |
| | | | | | L | 18 | |
| A 1775 | J1341+2622 | B1339+266B | 13 41 51.0 | + 26 22 23.0 | Ku | 38 | 1 |
| | | | | | C | 17 | |
| | | | | | L | 18 | |
| A 1795 | J1348+2633 | B1346+268B | 13 49 00.0 | +26 33 37.0 | Ku | 39 | 1 |
| | | | | | C | 26 | |
| | | | | | L | 26 | |
| A 2142 | J1558+2716 | B1556+274 | 15 58 14.0 | +27 16 20.0 | Ku | 23 | 3 |
| | | | | | C | 15 | |
| | | | | | L | 15 | |
| A 2255 | J1713+6407 | B1712+641 | 17 13 05.0 | +64 07 00.0 | Ku | 25 | 1 |
| | | | | | C | 15 | |
| | | | | | L | 15 | |
| A 2256 | J1703+7839 | C, B1706+787 | 17 03 28.0 | +78 39 55.0 | Ku | 12 | 4,1 |
| | | | | | C | 14 | |
| | J1700+7841 | I, B1703+787 | 17 00 51.0 | +78 41 25.0 | Ku | 11 | |
| | | | | | C | 14 | |
| Ophiuchus | J1712–2328 | D | 17 12 09.0 | –23 28 30.0 | Ku | 26 | 5 |

Col. 1: cluster name; Col. 2: source name according to J2000 IAU nomenclature; Col. 3: other source name according to reference given in Col. 8; Cols. 4 and 5: pointing position; Col. 6: observation band, see text; Col. 7: on-source integration time; Col. 8: references are 1 = Owen & Ledlow (1997), 2 = Vacca et al. (2014), 3 = Giovannini & Feretti (2000), 4 = Owen et al. (2014), and 5 = Murgia et al. (2010).

### 3.1. A 278

The tailed radio galaxy J0157+3214 in A 278 has been identified with the first-ranked galaxy of the cluster. Its total extent is about 2′, with the tail pointing toward the cluster center (Feretti et al . 1990, Owen & Ledlow 1997). The radio galaxy is displaced by about 2′ from the X-ray emission centroid, likely indicating that the cluster is at an early stage of evolution, in agreement with the tailed structure of the first-ranked galaxy. In the present observations at 15 GHz (map not shown here), the source is point-like at the resolution of 0.17″x 0.13″, with a flux density of 0.18 mJy.

### 3.2. A 520

This is a complex cluster that has been observed in the X-ray, optical, and radio domain (Vacca et. al 2014, and references therein). Govoni et al. (2001) revealed a wide radio halo whose largest linear extent is 1.4 Mpc, elongated in the NE-SW, as the location of the central X-ray emission. Deep Chandra observations indicate the presence of a bow shock in front of a dense clump to the southwest of the cluster (Govoni et al. 2004, Markevitch et al. 2005).

This cluster is characterized by two narrow NAT sources, with remarkably parallel tails oriented east, J0454+0255A and J0454+0255B, which are labeled D and E, respectively, by Vacca et al. (2014). The large-scale structure of these radio galaxies overlaid on the optical image from SDSS is shown in Fig. 1. The sources show total lengths of ∼ 500 and 300 kpc, respectively, they are located to the east of the cluster center, and both show tails oriented away from the cluster center. The two sources have been observed at 15 GHz with a single pointing.

The source J0454+0255A (Fig. 2) shows a nuclear structure with a symmetric extension and the evidence of a low brightness emission in the direction of the large-scale tail. The flux of the nucleus is 0.07 mJy.

The source J0454+0255B (Fig. 3) shows a 0.1 mJy nucleus and a knotty jet pointing to the opposite direction with respect to the large-scale tail, and then bending at less than 1″ from the core. By degrading the image to a lower resolution (not shown here), there is evidence of low brightness emission at the beginning of the tail.

### 3.3. A 869

This cluster is considered underluminous in X-ray band by Popesso et al. (2007), with an X-ray luminosity of $L_X \sim 2.41 \; 10^{43}$ (erg s$^{-1}$). The tailed radio galaxy J0946+0222, imaged by Owen & Ledlow (1997) at 20 cm with the VLA, shows a tail of more than 1′ in size. The images obtained here, presented in Fig. 4, show a one-sided and very colli-mated structure. At 1.4 GHz, the jet emanating from the core is very narrow at the beginning, then it widens to form the tail, which is easily visible for a large part of its total extent. At 5 GHz there is a hint of a bifurcation in the inner 1-2″ at the beginning of the tail, which might be an indication of twin faint jets.

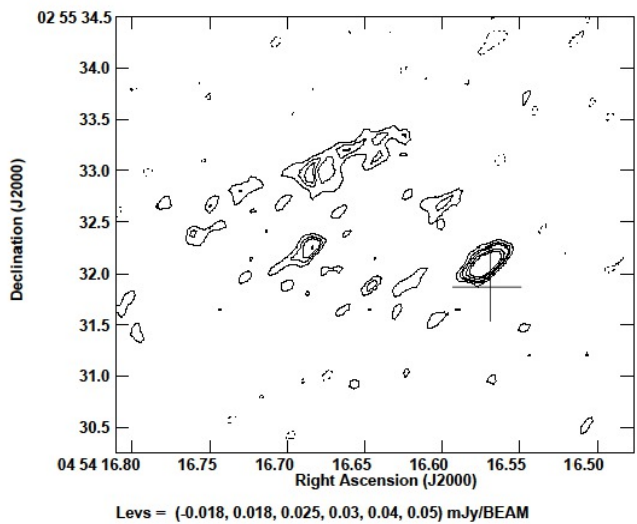

**Fig. 2.** Contour image of the radio galaxy J0454+0255A in A 520 at 15 GHz. The beam is $0.22'' \times 0.14''$ at -43°. The noise level is $6 \times 10^{-3}$ mJy/beam.

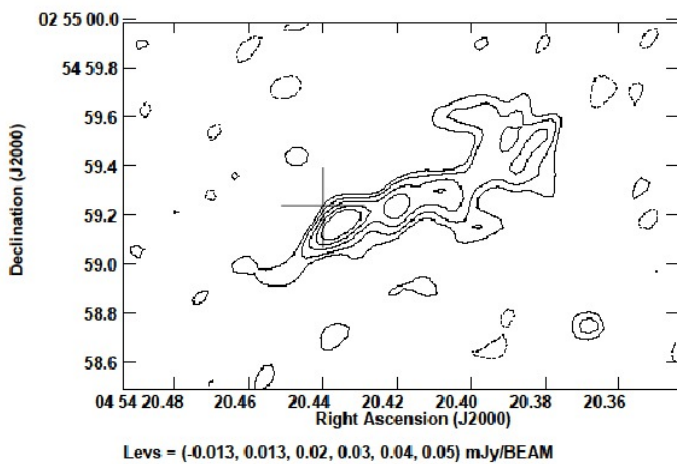

**Fig. 3.** Contour image of the radio galaxy J0454+0255B in A 520 at 15 GHz. The beam is $0.16'' \times 0.11''$ at -45°. The noise is $6.5 \times 10^{-3}$ mJy/beam.

In the image at 15 GHz (not shown here), only the nuclear source is detected. The nucleus has a flux of 15.8 mJy at 1.4 GHz, 11.5 mJy at 5 GHz, and 9 mJy at 15 GHz.

### 3.4. A 1775

The cluster A 1775 is a rich cluster hosting a binary galaxy system (dumb-bell galaxy) at its center (Giacintucci et al. 2007). The central cD galaxy is associated with a double radio galaxy not observed here, while its companion is the head-tail radio galaxy under study (J1341+2622). In large-scale images obtained by Owen & Ledlow (1997) and Giovannini & Feretti (2000), this radio galaxy shows a very narrow tail of about 8′ in extent.

In our image at 1.4 GHz, presented in the left panel of Fig. 5, the head-tail source J1341+2622 shows a total size of about 40″, and is transversally resolved. We note that the radio core is located south of the brightest spot in the L-band tail and in the 5 GHz tail, consistent with the galaxy position and also reported by Owen & Ledlow (1997). In the high-resolution image at 5 GHz (middle panel), the innermost structure of the source is detected as an off-center nucleus and two opposite symmetric jets. While the NE jet is rather straight in the direction of the tail, the SW jet shows a prominent bend at about 0.5″ (∼ 0.7 kpc) from the core. This structure is confirmed by the 15 GHz image (right panel), where the extended structure is mostly resolved out. The nucleus has a flux of 0.55 mJy at 5 GHz and 0.25 mJy at 15 GHz. The location of the optical galaxy is evident from the overlay of the 15 GHz radio images onto the optical image, shown in Fig. 6.

### 3.5. A 1795

A 1795 is a bright cool-core cluster, well observed at several wavelengths. It is characterized by a wealth of activity near the core, including core sloshing along with AGN feedback (Ehlert et al. 2015). The tailed radio galaxy J1348+2633, located at about 2.5′ from the cluster center, has been imaged by Owen & Ledlow (1997) and shows a total size of 45″. In our image at 1.4 GHz, shown in the top panel of Fig. 7, the nuclear region is slightly resolved and the jet/tail is very narrow and extended about 30″. At 5 GHz, only the nuclear region is detected with a central nucleus and two opposite jets in the E-W direction. The western jet widens at about 10″ to a very low brightness extended structure. At 15 GHz, the source reveals a nucleus of 0.14 mJy and a knotty jet toward the west. The eastern jet is detected at very low level of significance. From the comparison of the higher resolution images and the 1.4 GHz image, we deduce that the jets are bent on a scale of a few arcsec, to give rise to the tail.

### 3.6. A 2142

A 2142 is considered a cool-core cluster (see Govoni et al. 2010) because it shows a cool X-ray peak as well as strong evidence for a centrally enhanced metal abundance (e.g., White et al. 1994; Peres et al. 1998). On the other hand, optical features (Oegerle et al. 1995) and X-ray features, such as off-center small-scale structures and azimuthally asymmetric temperature variations, are indications of a merger (Henry & Briel 1996; Buote & Tsai 1996; Pierre & Stark 1998). Giovannini & Feretti (2000) confirmed the presence of a central diffuse emission in this galaxy cluster, similar to radio halos but of smaller size. A giant radio halo has recently been detected in this cluster (Farnsworth et al. 2013, Venturi et al. 2017). The tailed radio galaxy J1558+2716 is located at about 3′ from the cluster center. It is extended about 3′ and it is still one-sided at ∼ 1″ resolution at 3.6 cm and 6 cm (Govoni et al. 2010).

In our images, shown in Fig. 8, the tail is detected at 1.4 GHz for a total extent of ∼ 1′ and is very collimated.

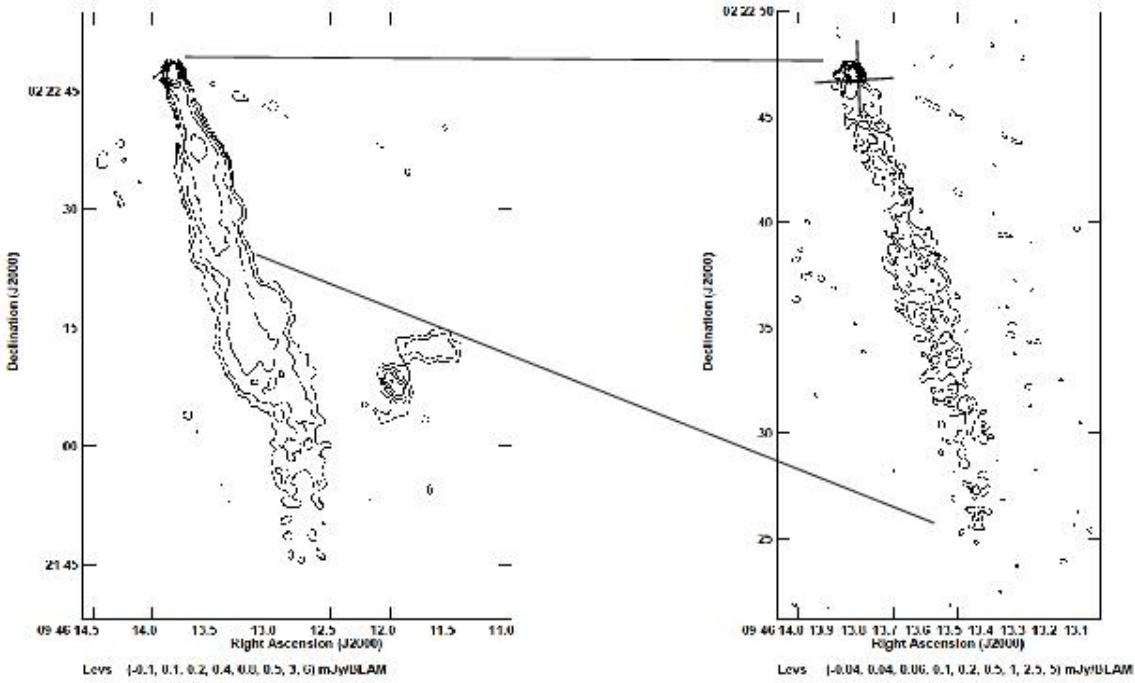

**Fig. 4.** Radio contour images of the tailed radio galaxy J0946+0222 in A869. *Left panel*: 1.4 GHz, with a beam of $1.28'' \times 1.15''$ at position angle $10°$. The noise is $3\ 10^{-2}$ mJy/beam. *Right panel*: 5 GHz, with a beam of $0.43'' \times 0.32''$ at $39°$. The noise is $1.2\ 10^{-2}$ mJy/beam.

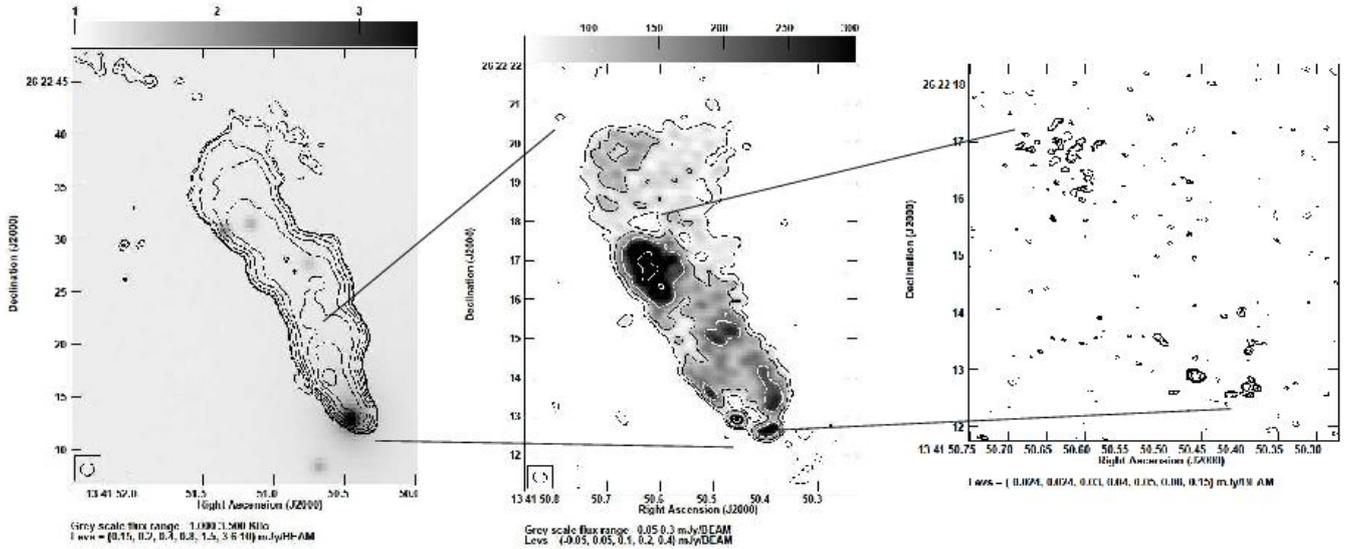

**Fig. 5.** Contour images of the radio galaxy J1341+2622 at the center of A 1775. *Left panel*: 1.4 GHz, with a beam of $1.29'' \times 1.18''$ at $-87°$, overlaid on the optical image from SDSS (gray scale). The noise in the radio image is $4.3\ 10^{-2}$ mJy/beam. *Middle panel*: 5 GHz with a beam of $0.35'' \times 0.27''$ at $80°$. The noise is $1.6\ 10^{-2}$ mJy/beam. *Right panel*: 15 GHz, with a beam of $0.19'' \times 0.11''$ at $76°$. The noise is $8\ 10^{-3}$ mJy/beam.

At 5 GHz, with the resolution of about $0.35''$, the nuclear structure is resolved in a core of 2.4 mJy and two well-separated bent jets. At the highest resolution of 15 GHz observations, the radio galaxy shows a bright nucleus of 1.7 mJy and two-sided symmetric jets, bent in the direction of the tail.

### 3.7. A 2255

Abell 2255 is a nearby rich galaxy cluster that has been extensively studied in several bands (e.g., Govoni et al. 2005, 2006, Pizzo et al. 2009). ROSAT X-ray observations indicate that A2255 has undergone a recent merger (Burns et al. 1995, Feretti et al. 1997, Davis et al. 2003). XMM-

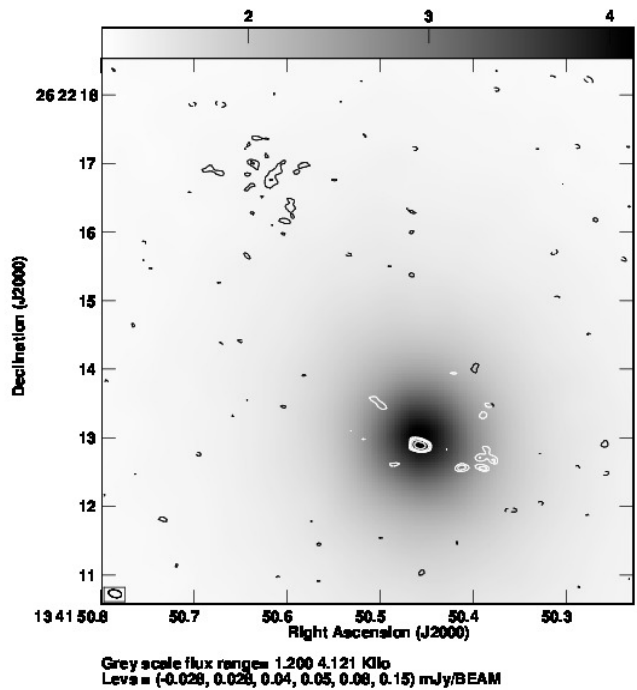

**Fig. 6.** Contour image of the radio galaxy J1341+2622 in A 1775 at 15 GHz overlaid on the gray-scale optical image from SDSS.

Newton observations revealed temperature asymmetries of the ICM suggesting that the merger occurred about 0.15 Gyr ago, probably along the E-W direction (Sakelliou & Ponman 2006). At radio wavelengths, A2255 shows a radio halo, external relics, and many embedded head-tail radio galaxies (Harris et al. 1980, Pizzo et al. 2009). In particular, our target is J1713+6407, (the Goldfish radio galaxy, so called by Harris et al. 1980), characterized by a prominent and slightly extended head and a very narrow tailed structure of about 2′ toward the south (Feretti et al. 1997). In our high-resolution image at 1.4 GHz, shown in the left panel of Fig. 9, the head is resolved and shows two opposite jets, the northern one characterized by a sharp bend toward east, in agreement with the image presented by Owen & Ledlow (1997). The innermost structure is easily detected at 5 GHz, with a compact core of 0.85 mJy and narrow collimated jets. In the image at 15 GHz at full resolution, only the nucleus of 1.05 mJy and the beginning of the northern jet are detected. We show here a 15 GHz image with degraded resolution, which reveals the northern jet up to $\sim$ 3″ from the core. A remarkable feature detected in both images, at 5 GHz and 15 GHz, is the gap in the northern jet at about 1.5″ from the core. A similar gap is also seen in the southern jet at 5 GHz and in the 1.4 GHz image of Owen & Ledlow (1997).

### 3.8. A 2256

The rich massive cluster A 2256 is characterized by an ongoing merger, which has been well studied in several bands, and in particular at radio wavelengths because it hosts a rich variety of radio phenomena (Bridle & Fomalont 1976; Bridle et al. 1979; Röttgering et al. 1994; Miller et al. 2003;

Clarke & Ensslin 2006; Kale & Dwarakanath 2010; van Weeren et al. 2012): a radio halo roughly coincident with the X-ray emission, a relic, several radio tails, a complex of very steep radio emissions, and more than 40 cluster members with detected radio emission. A complete radio study of this cluster has recently been presented by Owen et al. (2014). This clusters has two very narrow-angle tail radio galaxies, J1703+7839 and J1700+7841, labeled C and I, respectively, by Owen et al. (2014) and by previous authors.

The source J1703+7839, the so-called Long Tail radio galaxy, is one of the longest known tailed radio galaxies, with an extent of about 800 kpc. In the image obtained at 6 - 8 GHz with 0.3″ resolution, Owen et al. (2014) did not see any bifurcation near the core of possible twin bent-back tails and deduced that the Long Tail might be a one-sided jet. A similar conclusion was reached by the same authors for the source J1700+7841.

Our C-band image of J1703+7839, shown in the top panel of Fig. 10, shows the nucleus of 0.4 mJy and the tail for several arcsec. The high-resolution image at 15 GHz reveals a nuclear source of 0.14 mJy and the beginning of a one-sided jet, in agreement with the result of Owen et al. (2014). We note that the apparent one-sidedness could be due to sensitivity problems, given the weak detection of the primary jet.

The radio galaxy J1700+7841 shows on the large scale a tail toward the NE (see Fig. 11). The image at 5 GHz is presented in Fig. 12. The nucleus of 0.5 mJy is easily visible, with two opposite jets and some indication of bending in the western jet. From the comparison of the present image with the large-scale image, we can derive that the western jet is bent back by ram pressure. At 15 GHz (image not shown here) the source is unresolved, with a nucleus of 0.26 mJy.

### 3.9. Ophiuchus

Ophiuchus is one of the most massive and rich nearby galaxy clusters, located at about 12° from the Galactic center. It is one of the brightest clusters in X-ray band, the second after Perseus. Ophiuchus was originally believed to be a merging cluster, based on an X-ray temperature map derived from ASCA data and showing two very hot regions to west and south of the center (Watanabe et al. 2001). However, based on Suzaku data, these results were contradicted by Fujita et al. (2008), who did not detect the huge temperature variations found by Watanabe et al. (2001) and found that Ophiuchus is a cool-core cluster, with isothermal gas beyond 50 kpc from the cluster center. This is confirmed by the deep Chandra study of Werner et al. (2016) and by the detailed dynamical analysis of Durret et al. (2015), who derived that the cluster is very massive and old, characterized by very little star formation, and is therefore globally relaxed. Studies at radio wavelengths (Govoni et al. 2009, Murgia et al. 2010) revealed a mini-halo and several extended radio galaxies. The tailed radio galaxy J1712–2328 studied here is located in the SW cluster region, at about 7' from the cluster center, and is labeled D by Murgia et al. (2010).

The 15 GHz image of this radio galaxy overlaid onto the optical image is presented in Fig. 13. It shows twin jets, different in intensity and in transversal size. The overall structure is elongated to the south, following the structure of the eastern jet, which is rather straight at the beginning. The source core is likely identified with the eastern peak, which is slightly stronger than the western peak and is more compact, with a flux density of 0.14 mJy. The western peak is likely a bright knot in the western jet, which shows a prominent bend to the south, at few arcsec from the core, toward the low brightness tail.

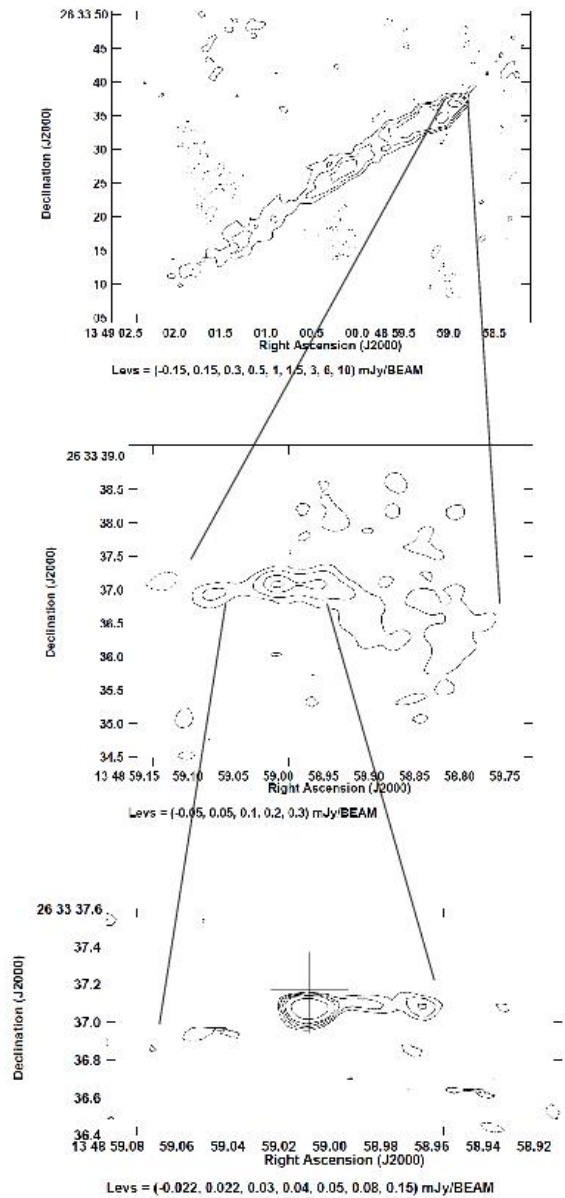

**Fig. 7.** Radio contours image of the radio galaxy J1348+2633 in A 1795. *Top panel*: 1.4 GHz, with a beam of $1.32'' \times 1.18''$ at $88°$. The noise is $6.8 \ 10^{-2}$ mJy/beam. *Middle panel*: 5 GHz, with a beam of $0.35'' \times 0.27''$ at $80°$. The noise is $1.6 \ 10^{-2}$ mJy/beam. *Bottom panel*: 15 GHz, with a beam of $0.19'' \times 0.11''$ at $76.5°$. The noise is $7.4 \ 10^{-3}$ mJy/beam.

## 4. Discussion

### 4.1. Structures

In Table 3 we summarize the source structures detected with the high-resolution images of the sources under study. Two-sided jets are present in all sources, except for J0157+3214 in A 278, where only the nuclear emission is detected, and J1703+7839 in A 2256, where only a one-sided jet aligned with the narrow tail is visible also in the Ku-band image (bottom panel of Fig. 10).

The jets detected in our images at the highest resolution are of low brightness, are resolved, and are not dominant with respect to the core emission. This morphology suggests that these jets are at most mildly relativistic on the kpc/sub-kpc scale. This is supported by the analysis of the ratio $R_{j-cj}$ between the jet and counter-jet brightness, which is related to the jet velocity as follows:

$$R_{j-cj} = \left(\frac{1 + \beta_j cos\theta}{1 - \beta_j cos\theta}\right)^{2+\alpha}, \quad (1)$$

where $\beta_j$ ($\beta_j = \frac{v}{c}$) is the jet velocity, and $\theta$ is the angle with respect to the line of sight, under the hypothesis that the jets are relativistic and intrinsically symmetric, thus the stronger jet is directed toward the observer and amplified by a Doppler-boosting effect.

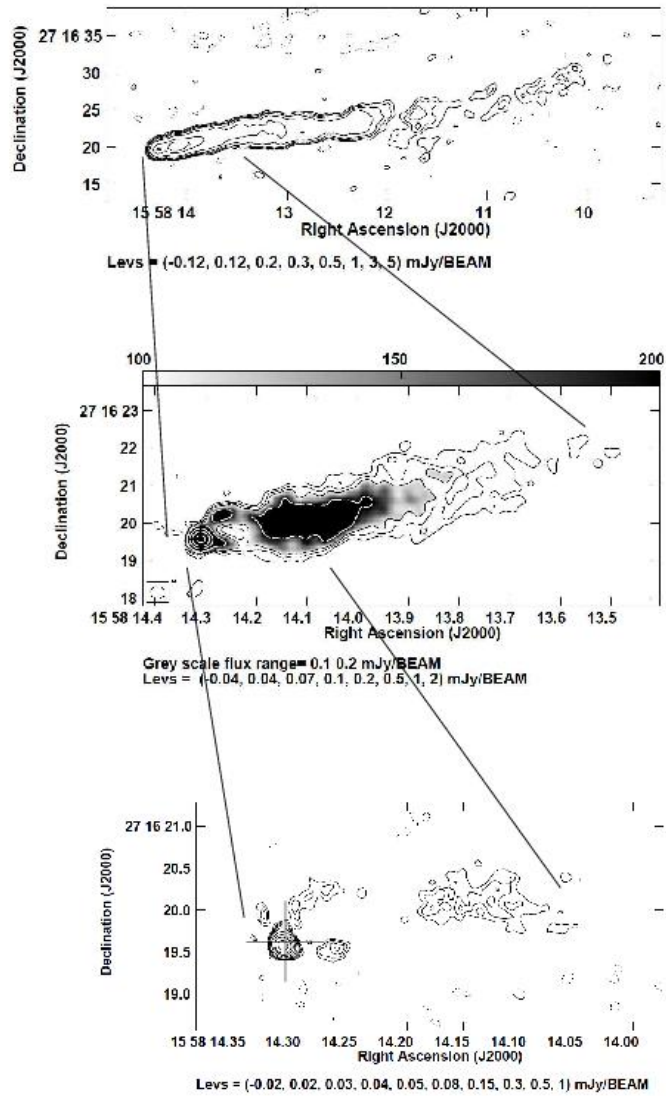

**Fig. 8.** Radio contours of the tailed radio galaxy J1558+2716 in A 2142. *Top panel*: 1.4 GHz, with a beam of $1.36'' \times 1.19''$ at $89°$. The noise is $4.3 \; 10^{-2}$ mJy/beam. *Middle panel*: 5 GHz, with a beam of $0.37'' \times 0.31''$ at $81°$. The noise is $1.3 \; 10^{-2}$ mJy/beam. *Bottom panel*: 15 GHz, with a beam of $0.16'' \times 0.13''$ at $55.5°$. The noise is $7 \; 10^{-3}$ mJy/beam.

**Table 3.** Summary of morphologies

| Cluster name | Source name | Source structure |
| --- | --- | --- |
| A 278    | J0157+3214  | no jet/tail detected |
| A 520    | J0454+0255A | C-shaped bent jet pair |
| A 520    | J0454+0255B | two-sided jets |
| A 869    | J0946+0222  | possible bifurcation at tail beginning |
| A 1775   | J1341+2622  | two-sided jets |
| A 1795   | J1348+2633  | two-sided jets |
| A 2142   | J1558+2716  | two-sided jets |
| A 2255   | J1713+6407  | two-sided jets |
| A 2256   | J1703+7839  | apparently one-sided jet |
| A 2256   | J1700+7841  | two-sided jets |
| Ophiuchus | J1712–2328 | two-sided jets |

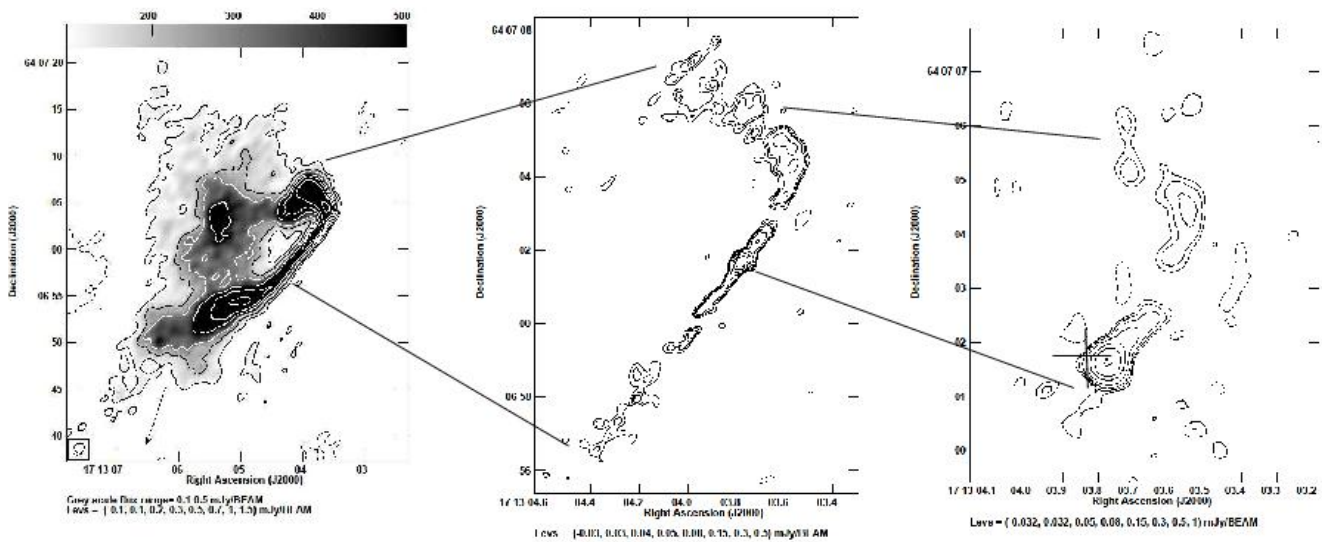

**Fig. 9.** Radio contours of the tailed radio galaxy J1713+6407 in A 2255. *Left panel*: 1.4 GHz, with a beam of 1.38″× 0.99″at -39° and a noise of 5 10$^{-2}$ mJy/beam. The arrow indicates the direction of the large-scale tail. *Middle panel*: 5 GHz, with a beam of 0.39″× 0.29″at -46° and a noise of 9 10$^{-3}$ mJy/beam. *Right panel*: 15 GHz, with a beam of 0.47″× 0.44″at -21° and a noise of 1.6 10$^{-2}$ mJy/beam.

We derived the values of the ratio between the jet and counter-jet brightness, when possible, from the Ku-band images. We present these values in Table 4, together with their estimated errors, and the implied values of $\beta_j \cos\theta$, obtained by assuming a value of 0.5 for the spectral index[1].

The above ratios are in most cases around 1.5, implying that jet velocities are of about 0.2c if the jet is oriented at about $60° - 70°$ with respect to the line of sight, and they are lower for more aligned jets (0.1c at $20° - 40°$). The sources J0454+0255B in A 520 and J1713+6407 in A 2255 shows slightly higher brightness ratios, which are still consistent with a mildly relativistic jet scenario, however. We conclude that no strong relativistic effects are involved in the jet morphologies of the sources studied here and that their properties are the same as those of the FRI sources.

We note that according to unified models, jets in FR I sources show a relativistic velocity near the core (see, e.g., Giovannini et al. 2001), but their velocity decreases within a few kpc, or even less, because of a strong interaction with the interstellar medium (Bicknell et al., 1990; Laing et al. 1999; see also the detailed study on 3C449 by Feretti et al. 1999).

### 4.2. Environment and bending

In order to investigate the environment of the radio galaxies under study, we searched in the literature for X-ray information about the cluster gas distribution by assuming that it can be approximated by a $\beta$-model (Cavaliere & Fusco-Femiano 1976),

$$n(r) = n_0 \left[1 + \left(\frac{r}{r_c}\right)^2\right]^{-\frac{3}{2}\beta}, \quad (2)$$

---
[1] $S(\nu) \propto \nu^{-\alpha}$

**Table 4.** Jet to counter-jet ratios

| Source name | $R_{j-cj}$ | $\sigma_{R_{j-cj}}$ | $\beta_j \cos\vartheta$ |
|---|---|---|---|
| J0454+0255B | 2.70 | 1.15 | 0.19 |
| J1341+2622 | 1.33 | 0.43 | 0.05 |
| J1348+2633 | 1.50 | 0.60 | 0.07 |
| J1558+2716 | 1.67 | 0.44 | 0.10 |
| J1713+6407 | $\gtrsim 3$ | | $\gtrsim 0.22$ |
| J1700+7841 | 1.20 | 0.16 | 0.4 |
| J1712−2328 | 1.60 | 0.20 | 0.09 |

where $n(r)$ is the density at a distance $r$ from the cluster center, $n_0$ is the central gas density, $r_c$ is the core radius of the gas density distribution, and $\beta$ is the ratio of the specific energy in galaxies to the specific energy in the hot gas. These parameters, scaled to the adopted cosmology, are given in Table 5, where we also list the values of the gas density $n_r$ reported at the radio galaxy location, assuming that its distance from the cluster center equals the projected distance. Therefore, since the true galaxy distance from the cluster center is unknown, the computed densities should be considered as upper limits. We derive that the density of the ICM at the radio galaxy location is very high, thus justifying the hypothesis that the bent structure is due to the interaction of high-velocity host galaxies moving through the dense ICM.

Following dynamical arguments, we used the expression derived from the Euler equation (Begelman et al. 1979, Jones & Owen 1979) to describe the jet bending conditions:

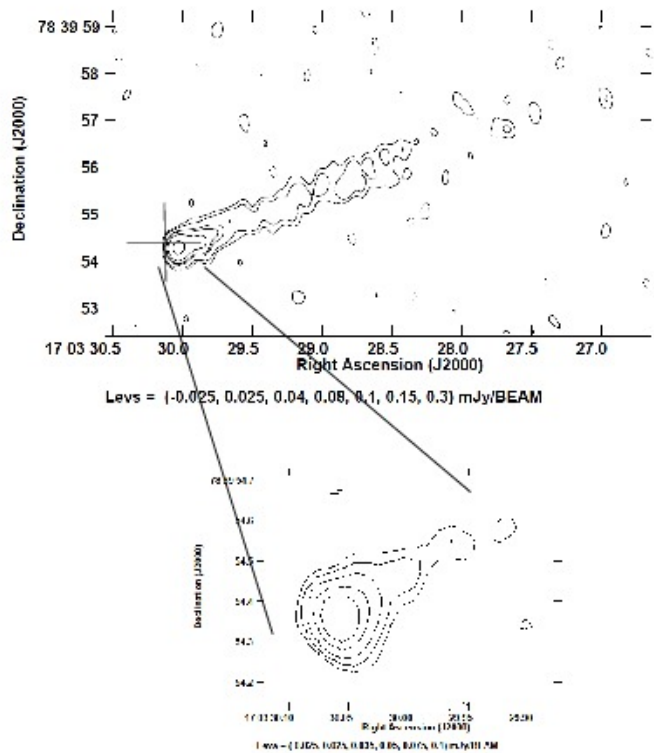

**Fig. 10.** Source J1703+7839 in A 2256. *Top panel*: 5 GHz at the resolution of $0.45'' \times 0.28''$ at $8°$. The noise in this image is $8 \; 10^{-3}$ mJy/beam. *Bottom panel*: 15 GHz at the resolution of $0.16'' \times 0.10''$ at position angle $32°$. The noise in this image is $1.2 \; 10^{-2}$ mJy/beam.

**Table 5.** Cluster properties

| Cluster name | Source name | $n_0$ $10^{-3}$ cm$^{-3}$ | $r_c$ kpc | $\beta$ | $r$ kpc | $n_r$ $10^{-3}$ cm$^{-3}$ | Notes | Ref |
|---|---|---|---|---|---|---|---|---|
| A 278 | J0157+3214 | 1.8 | 175 | 0.70 | 217 | 0.68 | merging | 1 |
| A 520 | J0454+0255A | 3.8 | 413 | 0.87 | 269 | 2.4 | merging | 2 |
|  | J0454+0255B |  |  |  | 358 | 1.8 |  |  |
| A 869 | J0946+0222 |  |  |  |  |  |  | no data |
| A 1775 | J1341+2622 | 3.3 | 122 | 0.58 | 58.5 | 2.7 | merging | 3 |
| A 1795 | J1348+2633 | 36 | 242 | 0.79 | 181 | 17 | cool core | 4 |
| A 2142 | J1558+2716 | 19 | 447 | 0.79 | 292 | 13 | merging/cool core | 4 |
| A 2255 | J1713+6407 | 2.1 | 411 | 0.79 | 350 | 1.1 | merging | 4 |
| A 2256 | J1703+7839 | 3.6 | 342 | 0.83 | 211 | 2.5 | merging | 4 |
|  | J1700+7841 |  |  |  | 596 | 0.65 |  |  |
| Ophiuchus | J1712−2328 | 19 | 187 | 0.70 | 212 | 8.2 | cool core | 4 |

Col. 1: cluster name; Col. 2: source name; Cols. 3, 4, 5: parameters of the cluster gas density distribution according to Eq. 2; Col. 6: projected distance of the radio galaxy from the cluster center; Col. 7: gas density at the radio galaxy location; Col. 8: note on the cluster dynamical status; Col. 9: reference to the cluster gas data (here scaled to the adopted cosmology), 1 = Feretti et al. (1990), 2 = Govoni et al. (2001), 3 = Cirimele et al. (1997), and 4 = Mohr et al. (1999).

$$R \sim h \left(\frac{\rho_j}{\rho_e}\right)\left(\frac{v_j}{v_g}\right)^2 \qquad (3)$$

where R is the jet radius of curvature, $\rho_j/\rho_e$ is the ratio between the jet density and the external density, $v_j/v_g$ is the ratio between the jet and galaxy velocity and $h$ is the scale height over which the ram pressure is transmitted to the jets.

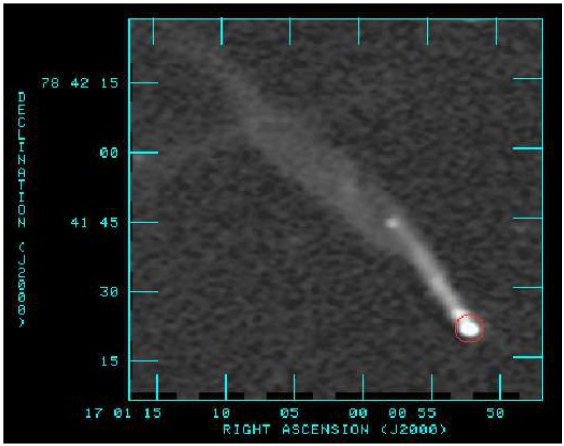

**Fig. 11.** Source J1700+7841 at 1.4 GHz on the large scale, from Owen et al. (2014).

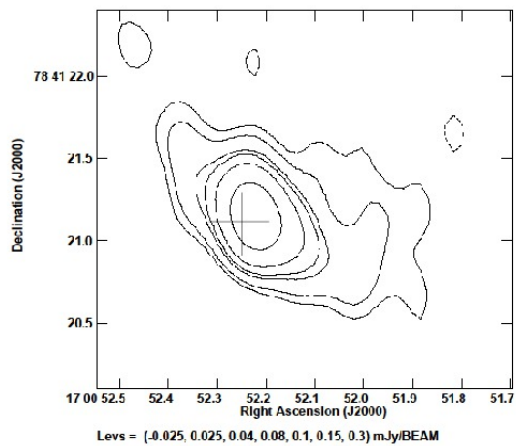

**Fig. 12.** Source J1700+7841 in A 2256 at 5 GHz at the resolution of $0.45'' \times 0.28''$ at a position angle of $10°$. The noise in this image is $8\ 10^{-3}$ mJy/beam.

For the galaxy velocity $v_g$, we assumed the dispersion of cluster galaxy velocities (from Wu et al. 1999), and for the jet velocity we assumed the same average value of 0.2c, which is derived to be consistent with the measured jet to counter-jet ratios. The jet density is unknown, therefore we consider two possibilities: a light jet with $(\rho_j/\rho_e) = 0.1$ (e.g., Clarke et al. 1986), and an extremely light jet with $(\rho_j/\rho_e) = 0.01$ (Massaglia et al. 2016). Moreover, we used the same value $h = 0.25$ kpc for each radio galaxy, which is estimated from the images and corresponds roughly to the transversal size of the jet. The latter assumption refers to the case of a jet that is directly in contact with the intergalactic medium, neglecting the possible shielding effect of the interstellar medium, which would lead to higher values of $h$. In this way, we obtain values of the jet curvature radius, which are given in Table 6. The calculated values of the curvature radius are indicative because of several assumptions, but they are much higher than those inferred from the radio images, except possibly for the source in A 2255. Therefore we deduce that projection effects are likely to play a significant role in the majority of our sources, that is, the observed jets/tails are very narrow because their plane is at a small angle to the line of sight.

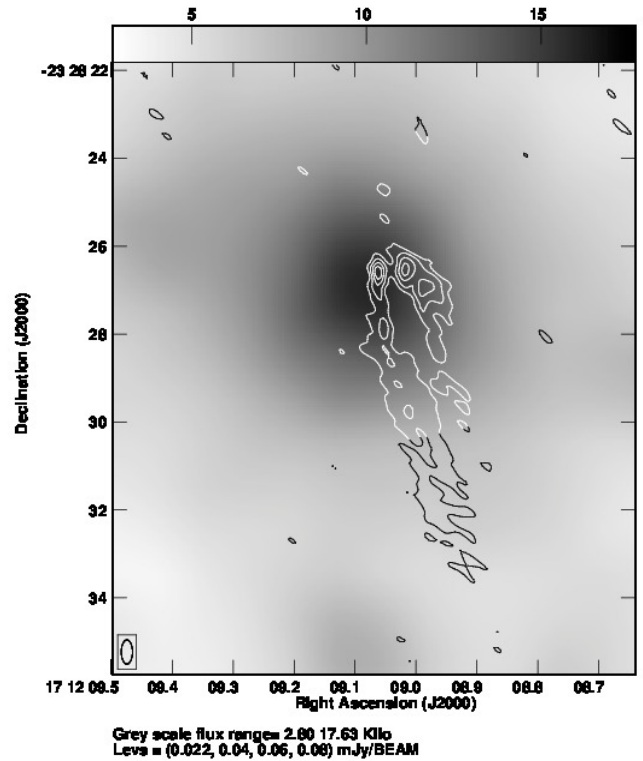

**Fig. 13.** Contour image of the tailed radio galaxy J1712–2328 in the Ophiuchus cluster at 15 GHz at a resolution of $0.5'' \times 0.25''$ at $0°$, overlaid on the gray-scale optical image from SDSS. The noise in this image is $7.1\ 10^{-3}$ mJy/beam.

**Table 6.** Results for the R curvature radius

| Source name | $v_g$ $km\ s^{-1}$ | $R_{calc0.1}$ kpc | $R_{calc0.01}$ kpc | $R_{obs}$ kpc |
|---|---|---|---|---|
| J0454+0255A | 988 | 92 | 9.2 | 2.2 |
| J1341+2622 | 1522 | 39 | 3.9 | 1.1 |
| J1558+2716 | 1132 | 70 | 7 | 0.6 |
| J1713+6407 | 1221 | 60 | 6 | 10 |
| J1712–2328 | 1050 | 82 | 8 | 0.4 |

Col. 1: source name; Col. 2: galaxy velocity; Col. 3: curvature radius calculated assuming $(\rho_j/\rho_e) = 0.1$; Col. 4: curvature radius calculated assuming $(\rho_j/\rho_e) = 0.01$; Col. 6 : curvature radius observed in the images.

### 4.3. Comparison with IC 310

IC 310 is peculiar because of its one-sided tail in the large structure and its one-sided pc-scale relativistic jet that has the same orientation. Another peculiar property of IC 310 is represented by its structure: in the 5 GHz image from the literature at a resolution of about $4''$, the source shows a strong core and an extended tail that is disconnected from the much brighter head (Dunn et al. 2010), with no indication of a physical connection between the two. This morphology strongly suggests a restarted activity of IC 310, where the low-brightness extended tail is likely the rem-

**Table 7.** Core and total powers

| Source name | z | $F_{core}$ mJy | $P_{core}$ W Hz$^{-1}$ | $F_{1.4GHz}$ mJy | $P_{1.4GHz}$ W Hz$^{-1}$ | Ref |
|---|---|---|---|---|---|---|
| J0157+3214  | 0.089 | 0.18 | 3.47 $10^{21}$ | 372  | 7.18 $10^{24}$ | 1 |
| J0454+0255A | 0.199 | 0.07 | 7.91 $10^{21}$ | 23.2 | 2.57 $10^{24}$ | 2 |
| J0454+0255B | 0.199 | 0.1  | 1.13 $10^{22}$ | 21.7 | 2.40 $10^{24}$ | 2 |
| J0946+0222  | 0.117 | 9.0  | 3.16 $10^{23}$ | 195  | 6.80 $10^{24}$ | 1 |
| J1341+2622  | 0.072 | 0.25 | 3.02 $10^{21}$ | 287  | 3.51 $10^{24}$ | 1 |
| J1348+2633  | 0.063 | 0.14 | 1.33 $10^{21}$ | 35   | 3.24 $10^{23}$ | 1 |
| J1558+2716  | 0.091 | 1.7  | 3.09 $10^{22}$ | 130  | 2.67 $10^{24}$ | 1 |
| J1713+6407  | 0.081 | 1.05 | 1.77 $10^{22}$ | 66   | 1.02 $10^{24}$ | 1 |
| J1703+7839  | 0.058 | 0.14 | 1.05 $10^{21}$ | 46   | 3.58 $10^{23}$ | 5 |
| J1700+7841  | 0.058 | 0.26 | 1.98 $10^{21}$ | 8.2  | 6.33 $10^{22}$ | 5 |
| J1712–2328  | 0.028 | 0.14 | 1.60 $10^{20}$ | 72   | 1.25 $10^{23}$ | 3 |
| IC 310      | 0.018 | 102  | 8.30 $10^{22}$ | 670  | 5.35 $10^{23}$ | 4 |

Col. 1: source name, Cols. 2: redshift, Col. 3: core flux density at 15 GHz from this paper for the present sources and from Piner & Edwards (2014) for IC 310; Col. 4: core power at 15 GHz; Col. 5: total flux density at 1.4 GHz from reference given in col. 7; Col. 6: total power at 1.4 GHz; Col. 7: references are 1 = Owen & Ledlow (1997), 2 = Vacca et al. (2014), 3 = Murgia et al. (2010), 4 = Ledlow & Owen (1995), and 5 = Röttgering et al. (1994).

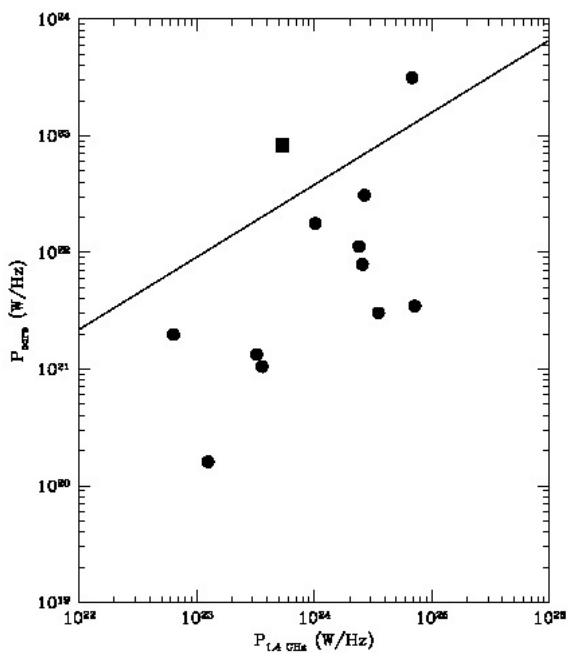

**Fig. 14.** Plot of the core power obtained from data at 15 GHz versus the total power at 1.4 GHz for the radio galaxies observed here and for IC 310 (square). Errors on powers are not indicated because they are as small as the symbols. The continuous line reports the correlation from the literature (see text).

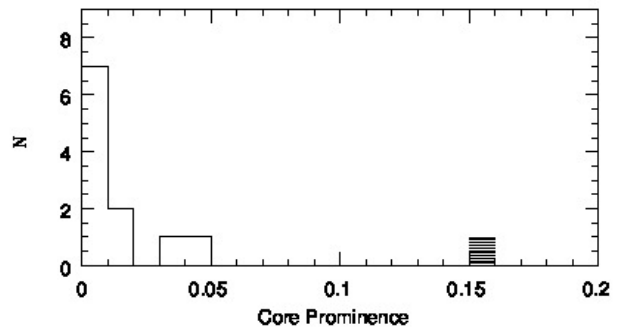

**Fig. 15.** Plot of the ratio of the core to total power displayed in Fig. 14. The radio galaxy IC 310 is shown as a dashed square.

nant of a past activity and is not affected by relativistic effects, while the bright core and the relativistic pc-scale jet are related to the restarted blazar-like activity. Moreover, the proper motion of galaxy IC 310 is aligned with the jet orientation, thus explaining the similarity between the relativistic jet orientation and the nonrelativistic extended tailed structure.

We found in Section 4.1 that the tailed radio galaxies of our sample are not affected by strong relativistic effects. To better investigate this point and to compare the properties of our sources with those of IC 310, we analyzed the core power obtained from our data at 15 GHz and the total power at 1.4 GHz obtained from the literature at low resolution. The data are given in Table 7 and are shown in the plot of Fig. 14, where the values of IC 310 are indicated as a full square.

Giovannini et al. (2001) and Giovannini (2004) showed that this comparison is a powerful indicator of the core

Doppler boosting and of the jet orientation and velocity. In Fig. 14 we plot as a continuous line the correlation found by Giovannini et al. (2001, 2004) between the total radio power at 408 MHz and the nuclear radio radio power at 5 GHz. The total radio power is scaled from 408 MHz to 1.4 GHz with a spectral index =1, while the core radio power is scaled from 5 GHz to 15 GHz with a spectral index =0.5.

We note that although significant scatter is present, most of our sources are below the correlation line because the nuclear radio power is not affected by relativistic amplification. Although the highest point in the figure refers to the radio galaxy J0946+0222 in A 869, the radio galaxy with the highest core to total power ratio is IC 310, as shown in Fig. 15. Thus IC 310 is characterized by the highest core prominence, consistent with its blazar nature. We also note that the core prominence of IC 310 is likely even higher when only the power of the restarted source is used as the total power.

More sensitive high-resolution images are needed to analyze the innermost structure of J0946+0222 in A 869. We would finally like to comment on the only one-sided source in our sample, that is, J1703+7839 in A 2256. The low core dominance and the faint one-sided jet suggest that relativistic effects are marginally present. This means that either the jet appears one-sided because of sensitivity problems, or the counter-jet has been disrupted by ram pressure, or an underlying twin-tail structure is not visible because of projection effects (see also Owen et al. 2014).

In the above scenario, IC 310 is a peculiar radio galaxy. All other NAT radio galaxies studied here are not different from classical NATs, but are strongly affected by projection effects.

## 5. Conclusions

We obtained subarcsecond-resolution images of head-tail galaxies with narrow tails that are unresolved on the arcsecond scale. We find that these sources are mostly characterized on the small scale by twin jets, which show sharp bends very close to the radio galaxy core.

From the analysis of the jet to counter jet ratio and from the study of the interaction between the jets and the ambient medium, we derive that these sources are not different from classical NATs, but are strongly affected by projection effects, or in other words, we conclude that these sources are seen mostly edge-on.

From the plot of the core power versus total power of the radio galaxies under study and of IC 310, we derive that the core prominence in our sources is low, thus confirming that they are genuine FRI sources. Our data demonstrate that high-resolution observations are necessary to properly analyze very detailed structures close to the core.

*Acknowledgements.* The National Radio Astronomy Observatory is a facility of the National Science Foundation operated under cooperative agreement by Associated Universities, Inc. We thank the anonymous referee for constructive comments that significantly helped to improve the data presentation and the discussion.